Influence of Stretching Boundary Conditions on Fracture in Phantom Star Polymer Networks: From Volume to Cross-sectional Area Conservation

*Yuichi Masubuchi, Takato Ishida, Yusuke Koide, and Takashi Uneyama

Department of Materials Physics, Nagoya University, Nagoya 4649603, Japan

*To whom correspondence should be addressed mas@mp.pse.nagoya-u.ac.jp

Ver Sep 2, 2025

ABSTRACT

This study systematically investigates the effect of stretching boundary conditions—ranging from conservation of cross-sectional area to conservation of volume—on the rupture behavior of phantom star polymer networks using energy-minimizing coarse-grained molecular simulations. By continuously varying the deformation parameter, the simulations reveal that true stress and rupture characteristics, such as strain and stress at break and work for rupture, systematically decrease as the boundary condition approaches cross-sectional area conservation. In contrast, nominal stress and the corresponding rupture characteristics exhibit near-independence from boundary conditions, indicating that bond tension remains largely unaffected for phantom networks under the examined conditions. These results clarify that volume expansion primarily drives deviations in true stress and highlight a critical distinction between true and nominal stress-strain definitions. The difference between true and nominal stress-strain relations also affected the scaling exponent for strand length dependence on stretch at break. The findings stress the importance of specifying both deformation boundary conditions and stress-strain definitions in polymer network simulations for accurate interpretation of mechanical properties.

INTRODUCTOIN

The mechanical fracture of polymer networks is a critical topic in polymer science, with significant implications for the design of gels, rubbers, and a wide range of soft materials [1,2]. Over the past decades, molecular simulations have played a pivotal role in unraveling the fracture mechanisms of these complex networks, as they can reveal microscopic details inaccessible to experiment [3]. Foundational simulation studies, such as those by Stevens [4,5], introduced coarse-grained bead-spring models [6,7] to represent densely cross-linked epoxy-like networks, providing early insights into the impact of wall anchoring and interfacial properties on mechanical failure. Subsequent research expanded this approach, examining how parameters such as crosslinker functionality [8,9], polymer rigidity [10], ionic interactions [11], and topological entanglements [12] influence fracture behaviors.

A major advancement in the field was the transition from wall-bound simulations to free boundary conditions, enabling more realistic modeling of glassy and elastomeric polymers and their fracture dynamics [13–15]. For example, Panico et al. [16] systematically investigated glassy polymer networks under alternative boundary conditions, thereby revealing the influence of network architecture on mechanical response. Such efforts have elucidated the roles of chain rigidity [17], entanglements [18,19], dangling ends and loops [20,21], bimodal strand lengths [22,23], node functionality [24–27], prepolymer concentration [28], strand molecular weight [29], and cross-linking inhomogeneity [30] across an array of polymer systems, including nanocomposites [31], double networks [32], slide-ring networks [33], and fully atomistic models [34–36].

Despite this progress, the impact of stretching boundary conditions—that is, the deformation protocols imposed during uniaxial tension—remains relatively underexplored. Most prior simulations employing solid walls kept the cross-sectional area orthogonal to the stretching direction fixed [4,5,8–12]. In other cases, including studies without explicit walls [32–34], a fixed area was likewise enforced, often to focus on local phenomena such as crack propagation. Alternatively, some researchers have maintained constant lateral pressure or conserved system volume during stretching [16–18,21,30,35,37]. Experimental Poisson's ratios have occasionally been used to deform simulation boxes [31], while other studies assume volume conservation by constricting the transverse dimensions during elongation [19,20,22–29].

Critically, however, there has been no systematic comparison of different deformation protocols, nor an analysis of how the choice of boundary condition affects fundamental rupture criteria. For example, fixing the cross-section can artificially promote bond overstretching due to volumetric expansion, while volume-conserving protocols may better reflect experimental setups but influence stress distribution differently. These nuances become especially important at high extensions or near the limits of bond strength, where the coupling between deformation modes is non-negligible.

In this context, the present study aims to address this gap by systematically investigating the effects of stretching boundary conditions on the rupture of phantom star polymer networks using coarse-grained simulations. By introducing a deformation parameter $\varsigma$, analogous to Poisson's ratio, we interpolate between the boundary conditions of cross-sectional area conservation ($\varsigma =$

0) and volume conservation ($\varsigma = 0.5$). We analyze how true and nominal stress-strain relations, as well as rupture characteristics and scaling behaviors, depend on the chosen deformation protocol. The findings provide concrete guidance for interpreting both computational and experimental results and underscore the need for explicit reporting of stress-strain definitions and boundary conditions in studies of polymer network mechanics.

MODEL AND SIMULATIONS

Coarse-grained molecular simulations were employed to investigate the fracture behavior of phantom star polymer networks under varying deformation conditions, closely following methodologies established in prior work [24–29]. The simulation protocol centers on bead-spring star branch prepolymers, which were uniformly dispersed within a cubic periodic box at a fixed bead number density $\rho$. Each star polymer features a monodisperse architecture characterized by an equal number of arms $f$ and identical bead count per arm $N_a$.

Inter-bead interactions are limited exclusively to bonded connectivity between consecutive beads, omitting excluded volume and attractive forces to maintain the phantom nature of the network. Following thorough thermal equilibration, network formation (gelation) proceeds via a Brownian dynamics-driven end-linking reaction, with reactive encounters restricted to different polymers to suppress the formation of trivial loops. This gelation protocol is designed to mimic experimental procedures for tetra-PEG gels [38]. Network snapshots are recorded at various conversion rates $p$. Cycle rank density $\xi$, the number of minimal topological circuits per branch point, is analyzed to confirm consistency with mean-field predictions for random cross-linking [39].

After gelation, the system undergoes energy minimization, alternated with incremental, affine elongation steps until mechanical rupture occurs. The deformation protocol is described by the deformation rate tensor:

$$\boldsymbol{\kappa} = \begin{pmatrix} \dot{\varepsilon} & 0 & 0 \\ 0 & -\varsigma\dot{\varepsilon} & 0 \\ 0 & 0 & -\varsigma\dot{\varepsilon} \end{pmatrix} \quad (1)$$

where $\dot{\varepsilon}$ represents the elongation rate at each elongation increment, and is related to the simulation box size in the elongational direction $L$ as $\varepsilon = \ln L/L_0$, with its initial value $L_0$. The deformation parameter $\varsigma$, which interpolates boundary conditions from cross-sectional area conservation to volume conservation, is defined as follows, similarly to Poisson's ratio [40];

$$\varsigma = -\frac{\varepsilon_t}{\varepsilon} = -\frac{\ln L_t/L_0}{\ln L/L_0} \quad (2)$$

Here, $\varepsilon_t$ and $L_t$ are the true (Hencky) strain and the simulation box size for the transverse directions.

The total potential energy is given by

$$U = -\frac{3k_B T b_{\max}^2}{2a^2} \sum_{i,k} \ln\left(1 - \frac{\mathbf{b}_{ik}^2}{b_{\max}^2}\right) \quad (3)$$

where $k_B T$ is the thermal energy, $\mathbf{b}_{ik}$ is the vector between $i$-th and $k$-th beads, $a$ is the equilibrium bond length, and $b_{\max}$ is the non-linear parameter for bond stretch. Brownian dynamics utilize forces derived from this potential, while post-gelation energy minimization is performed using the Broyden-Fletcher-Goldfarb-Shanno (BFGS) optimization algorithm [41].

After each elongation-minimization cycle, all bond vectors are monitored. If the maximum bond length exceeds the critical length $b_c = 0.9 b_{\max}$, the corresponding bond is removed, and the energy minimization is repeated, until the maximum bond length becomes below $b_c$, without stretch. This sequential scheme enhances computational efficiency relative to previous fragmentation protocols [24–28], and yields results consistent with bulk network properties [29].

All simulations are executed using nondimensional units, with intrinsic length, energy, and time scales defined as $a$, $k_B T$, and $\tau = \zeta a^2 / k_B T$, where $\zeta$ is the friction coefficient applied during Brownian motion. Typical system sizes contain approximately 35,000 beads, with minor adjustments made for variants in $f$ and $N_a$. The integration time step is set to $\Delta t = 0.01$, while the reaction cutoff $r_c = 0.5$ and cumulative reaction rate $p_r = 0.1$ remain fixed as previously described [24]. Network elongation is incremented with steps of $\dot\varepsilon = 0.1$, energy minimization proceeds with step size $\Delta r = 10^{-2}$, and the conversion parameter for optimization is $\Delta u = 10^{-4}$.

Simulations span a parameter space, with the deformation parameter $\varsigma$ ranging from 0 to 0.5 for $N_a = 5$, and further series for $N_a = 2$ to 16 at $\varsigma = 0$ and 0.5. The branching arm number $f$ varies from 3 to 8, and prepolymer conversion $p$ ranges from 0.6 to 0.95. System size dependence and boundary effects have previously been established [24] as negligible for fracture results within this modeling framework.

RESULTS AND DISCUSSION

Typical snapshots of simulated networks under elongation are shown in Figure 1 for the deformation parameter $\varsigma$ at 0 and 0.5, with node functionality $f = 4$, conversion $p = 0.95$, and prepolymer arm length $N_a = 5$. As expected, the total simulation box volume increases when $\varsigma = 0$ (cross-sectional area conserved), whereas it remains constant for $\varsigma = 0.5$ (volume conserved). At high strains, structural inhomogeneities are observed; however, their resemblance across both boundary conditions suggests they arise primarily from force balance rather than from volumetric effects. The critical Hencky strain at rupture ($\varepsilon_b = 2.69$) remains unaffected by the choice of boundary condition, for this specific initial condition.

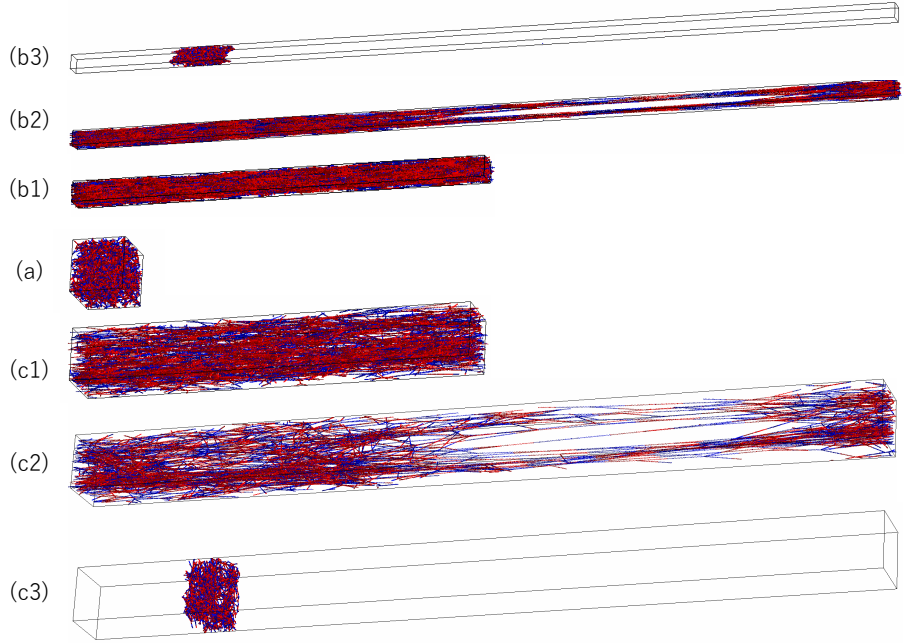

**Figure 1** Typical Snapshots for the networks with $f = 4$, $p = 0.95$, and $N_a = 5$. The structure after energy minimization before stretch is shown in (a). Panels (b) and (c) are for the cases with $\varsigma = 0.5$ and 0. The applied Hencky strains are 2.0 for (b1) and (c1), 2.68 for (b2) and (c2), and 2.69 for (b3) and (c3). Segments in blue and red indicate different chemistries, only between which the end-linking reaction is allowed.

From a given snapshot, the stress tensor is calculated by Kramers standard formula [42].

$$\sigma_{\alpha\beta} = \frac{1}{V} \sum_i f_{i\alpha} b_{i\beta} \qquad (4)$$

Here, $V$ is the volume, and $\mathbf{f}_i$ is the tension acting on bond $i$. From the stress tensor, true and nominal elongational stress, $\sigma_E$ and $\sigma_{En}$, are obtained, as written below.

$$\sigma_E = \sigma_{xx} - \varsigma(\sigma_{yy} + \sigma_{zz}) \qquad (5)$$
$$\sigma_{En} = \sigma_E A/A_0 = \sigma_E \exp(-2\varsigma\varepsilon) \qquad (6)$$

The subscripts $x$, $y$, and $z$ refer to the stretching and perpendicular directions. $A$ is the cross-sectional area normal to the stretching direction, and $A_0$ is its initial value. For work conjugacy, $\sigma_E$ and $\sigma_{En}$ are plotted against true and nominal strain, $\varepsilon$ and $\varepsilon_n$:

$$\lambda = \frac{L}{L_0} \qquad (7)$$

$$\varepsilon = \ln \lambda \qquad (8)$$
$$\varepsilon_n = \lambda - 1 \qquad (9)$$

Figure 2 presents true and nominal stress–strain curves for networks stretched under various deformation parameters. The true stress $\sigma_E$ plotted against true strain $\varepsilon$ reveals systematic changes: both the stress at break and modulus decrease as $\varsigma$ decreases, corresponding to an increasingly larger volume. In addition, the sharpness of the peak at rupture is progressively lost at lower $\varsigma$, reflecting the dominant effect of volumetric change rather than increasing bond tension.

In contrast, nominal stress $\sigma_{En}$ plotted against nominal strain $\varepsilon_n$ displays negligible dependence on $\varsigma$. The overlap between curves for different boundary conditions demonstrates that, for these phantom networks, the tension acting on each bond is essentially insensitive to the overall deformation protocol when analyzed in the nominal frame. In other words, the bond stretch to transverse directions does not significantly contribute to the results. The dullness of rupture peaks in nominal stress–strain plots is due to neglecting the reduction in cross-sectional area for cases with $\varsigma > 0$.

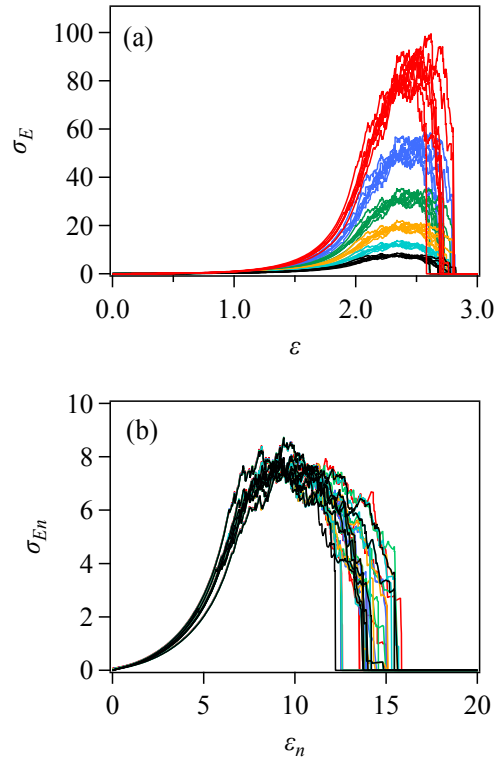

**Figure 2** Stress-strain curves for $f = 4$, $p = 0.95$, $N_a = 5$, and deformation parameter $\varsigma = 0$

(black), 0.1 (sky-blue), 0.2 (orange), 0.3 (green), 0.4 (blue), and 0.5 (red) for true stress versus true (Hencky) strain (a) and nominal stress versus nominal strain (b). Each curve shows an individual simulation run.

From stress-strain curves, characteristic quantities were extracted: strain at break $\varepsilon_b$, stress at break $\sigma_b$, and work for rupture $W_b$. Figure 3 shows these quantities for varying node functionalities ($f = 3\sim8$) at $p = 0.95$ and $N_a = 5$ as a function of $\varsigma$, using both true and nominal stress–strain definitions. When calculated from true stress/strain ($\sigma_E$ vs $\varepsilon$), all characteristic values increase monotonically with $\varsigma$, reflecting the mechanical influence of volume change (left column). Conversely, rupture values extracted from nominal stress/strain ($\sigma_{En}$ vs $\varepsilon_n$) remain effectively constant across all deformation protocols (right column).

One may argue that since the stretch at break $\lambda_b$ should be identical irrespective of the stress-strain definitions, the different $\varsigma$−dependence between $\varepsilon_b$ and $\varepsilon_{nb}$ is unconvincing. This result is due to the definition of rupture, which is based on the peak position in the stress-strain curve. Namely, the stretch corresponding to the peak position is slightly larger for true stress/strain definition than that for nominal stress/strain one due to the change of cross-section. This peak shift is more significant for a larger $\varsigma$ value.

A detailed explanation would also be warranted for $W_b$. For incompressible systems, the work density (energy density), defined as the integral of the stress–strain curve, should theoretically yield the same value whether nominal or Hencky strain is adopted. However, the observed discrepancy in work density between these two definitions is not paradoxical—rather, it reflects a difference in physical interpretation. In the Hencky strain formalism, $\sigma_E d\varepsilon$ represents the infinitesimal work input normalized by the current volume, yielding an energy density referenced to the deformed state. Conversely, the nominal strain approach interprets $\sigma_{En} d\varepsilon_n$ as the infinitesimal work input normalized by the initial volume, referencing the undeformed state. When the system volume changes during deformation, these interpretations naturally diverge. If constituents in the system are conserved, the work density calculated by nominal strain is proportional to the average work per constituent, which matches the intuitive physical meaning of work density. In contrast, in the Hencky strain framework, changes in system size alter the constituent density, leading to an apparent dilution of the system and a decrease in work density as the volume increases.

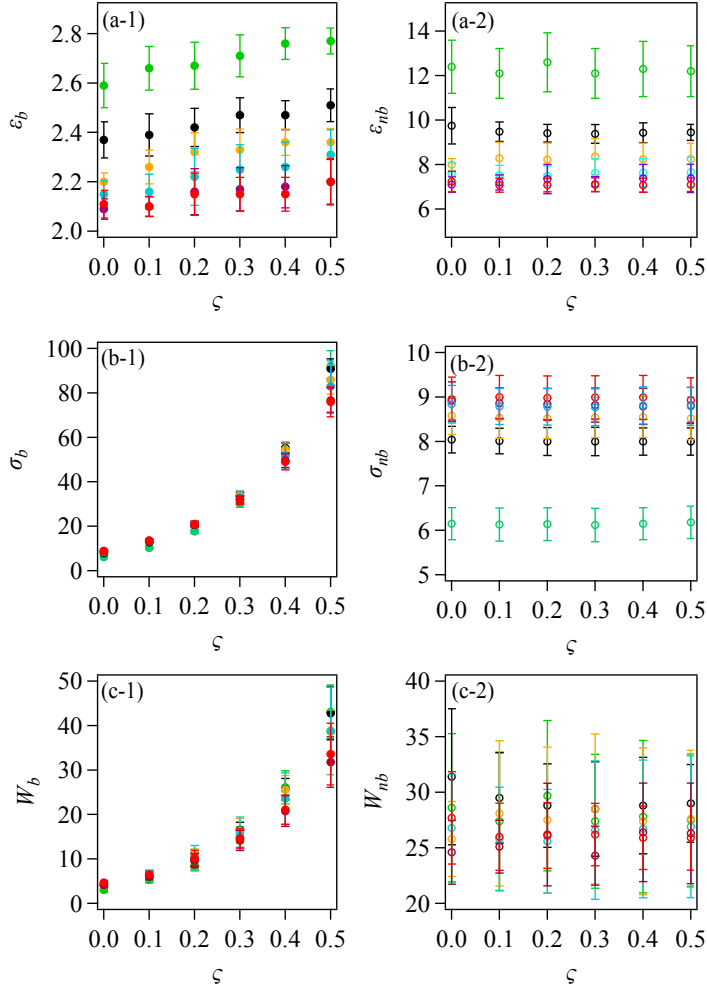

**Figure 3** Strain and stress at break $\varepsilon_b$ (a) and $\sigma_b$ (b), and work for rupture $W_b$ (c) plotted against the deformation parameter $\varsigma$ for $N_a = 5$, and various node functionalities $f = 3$ (green), 4 (black), 5 (orange), 6 (sky-blue), 7 (violet), and 8 (red). The left column (column 1) shows the values obtained from the stress-strain relation between true stress and true strain, whereas the right column (column 2) is for nominal stress and nominal strain. The error bar shows the standard deviation among eight independent simulation runs at each condition.

Figure 4 explores rupture characteristics as a function of cycle rank density $\xi$. This topological parameter that quantifies network circuits relative to branch point density, for multiple values of $f$ and $p$ [25]. As shown, master curves emerge for all data (when normalized by branch point density $\varphi_{br}$ for $\sigma_b$ and $W_b$), and exhibit power-law behavior for both true and nominal rupture quantities. For true stress/strain, the apparent power-law exponent slightly increases as $\varsigma$ decreases due to volume increases at high stretch. In contrast, nominal stress/strain master curves display no such sensitivity to boundary condition, emphasizing the relevance of the stress–strain definition to network analysis.

Concerning $\varepsilon_b$ and $\varepsilon_{nb}$ in Fig. 4 (a-1) and (a-2), it seems implausible that both quantities independently follow a power-law relationship, despite their definitions given by eqs 8 and 9. Indeed, $\varepsilon_{nb}$ exhibits a slight concave curve in a large $\xi$ range. Nevertheless, the difference is concealed by scattering and errors in the data.

Concerning $\sigma_{nb}$, the result demonstrates that $\sigma_{nb}(\xi)$ is independent of $\varsigma$, suggesting that under large extension, subchains deform predominantly along the stretching direction, largely independent of $\varsigma$. This interpretation is consistent with the behavior illustrated in Figure 1. Since the coupling between the stretching and transverse directions could depend on $\xi$, this result may reflect a universality inherent to the class of examined random networks.

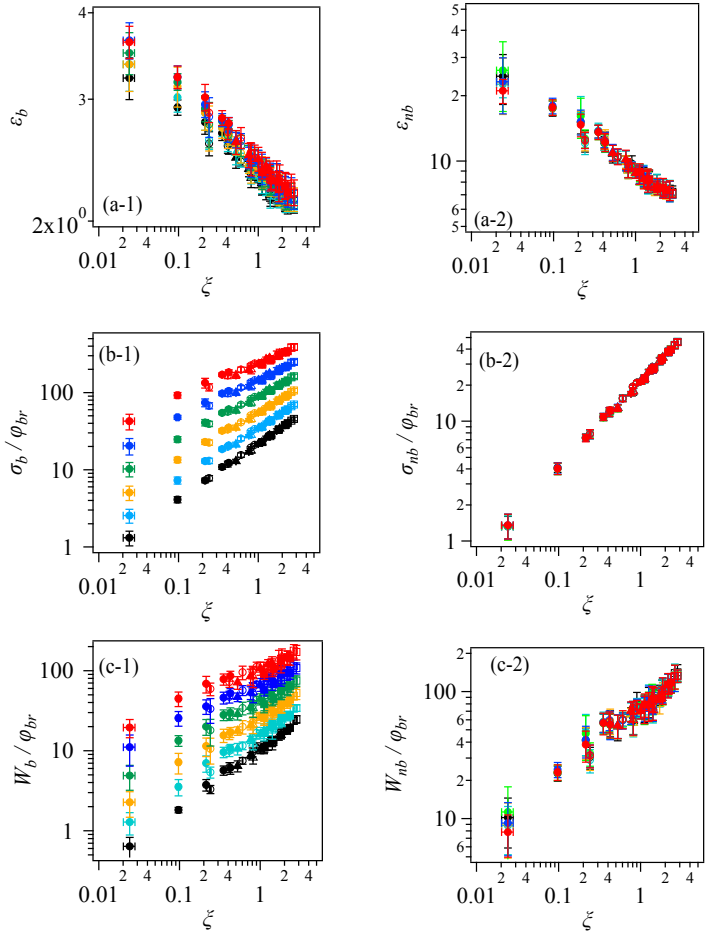

**Figure 4** Strain and stress at break, $\varepsilon_b$ (a) and $\sigma_b$ (b), and work for rupture $W_b$ (c) plotted

against cycle rank density $\xi$ for $N_a = 5$ with various $f$, $p$, and $\varsigma$. Panels in column 1 on the left show the values obtained from $\sigma_E - \varepsilon$ curves, whereas those in column 2 on the right are for $\sigma_{En} - \varepsilon_n$ plots. $\sigma_b$ and $W_b$ are normalized by the branch point density $\varphi_{br}$ before stretch. Symbols indicate $f = 3$ (filled circle), 4 (unfilled circle), 5 (filled triangle), 6 (unfilled triangle), 7 (filled square), and 8 (unfilled square). Colors indicate different $\varsigma$ values as 0 (black), 0.1 (sky-blue), 0.2 (orange), 0.3 (green), 0.4 (blue), and 0.5 (red). Error bars correspond to the standard deviations for eight different simulation runs for each condition.

Further insight is gained by examining the scaling of stretch at break $\lambda_b$ with strand length $N_s = 2N_a + 1$ for distinct deformation conditions shown in Fig. 5. When derived from true stress/strain with $\varsigma = 0.5$, a relation $\lambda_b \sim N_s^{0.66}$ is observed as reported previously [29], consistent with experimental results for tetra-PEG gels [43]. However, analysis of nominal stress/strain consistently yields the more widely accepted relation $\lambda_{nb} \sim N_s^{0.5}$, in agreement with single-strand models [44,45]. For $\varsigma = 0$, both definitions coincide due to the equivalence of true and nominal stress, and give $\lambda_n \sim N_s^{0.5}$.

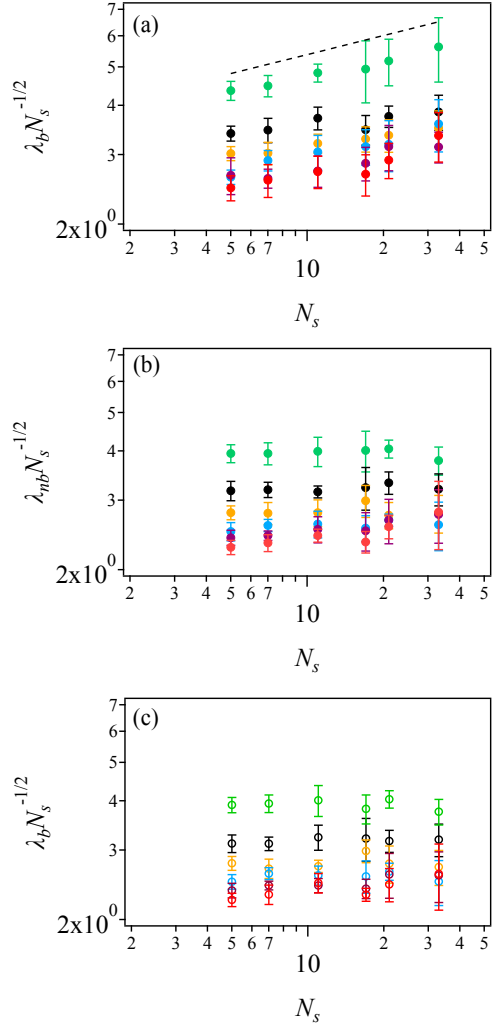

**Figure 5** Stretch at break $\lambda_b$ plotted against the strand molecular weight $N_s$ normalized by $N_s^{1/2}$ with $\varsigma = 0.5$ for $\sigma_E - \varepsilon$ curves (a), $\varsigma = 0.5$ for $\sigma_{En} - \varepsilon_n$ curves (b), and with $\varsigma = 0$ (c). $f = 3$ (green), 4 (black), 5 (orange), 6 (sky-blue), 7 (violet), and 8 (red). $p = 0.95$. Error bars correspond to the standard deviations for eight different simulation runs for each condition. The broken line in panel a shows a slope of 0.16.

## CONCLUSIONS

This study systematically explored the influence of deformation boundary conditions on the rupture behavior of phantom star polymer networks using energy-minimizing molecular simulations. By varying the deformation parameter $\varsigma$, which governs the transverse-to-elongational strain ratio and spans from conservation of cross-sectional area ($\varsigma = 0$) to conservation of volume ($\varsigma = 0.5$), we discussed insights into how stretching protocols affect network fracture characteristics.

The relationship between true stress and true strain was strongly sensitive to ς. Specifically, as the deformation condition approached cross-sectional area conservation, rupture-related parameters—including strain at break, stress at break, and work for rupture—systematically decreased. This ς-dependence arises primarily from volume changes associated with different boundary conditions. In contrast, nominal stress and nominal strain analyses showed negligible sensitivity to ς, indicating that bond tensions remain virtually unchanged under the studied deformation protocols in phantom networks. The distinction between true and nominal stress-strain frameworks was also shown to impact the interpretation of strand molecular weight dependence on stretch at break, modifying the scaling exponent. These results collectively underscore the critical importance of explicitly specifying deformation boundary conditions and adopting consistent stress-strain definitions in polymer network simulations to ensure accurate mechanical characterization.

Given that this work focuses on phantom networks without excluded volume or attractive interactions, future investigations are warranted to examine how these inter-segment interactions may further modulate rupture properties under different deformation conditions. Such studies are underway.

ACKNOWLEDGEMENTS
The Hibi Science Foundation partly financially supported this work.